\documentclass[journal]{IEEEtran}
\usepackage{siunitx}
\usepackage{adjustbox}
\usepackage{makecell}
\usepackage{xcolor}

\usepackage{booktabs} % for better horizontal lines

\ifCLASSINFOpdf
  \usepackage[]{graphicx}
\else
  \usepackage[dvips]{graphicx}
\fi
\begin{document}
%
% paper title
% Titles are generally capitalized except for words such as a, an, and, as,
% at, but, by, for, in, nor, of, on, or, the, to and up, which are usually
% not capitalized unless they are the first or last word of the title.
% Linebreaks \\ can be used within to get better formatting as desired.
% Do not put math or special symbols in the title.
\title{Scalable 49-Channel Neural Recorder with an Event-Driven Ramp ADC and PCA Compression in 28~nm CMOS}
%
%
% author names and IEEE memberships
% note positions of commas and nonbreaking spaces ( ~ ) LaTeX will not break
% a structure at a ~ so this keeps an author's name from being broken across
% two lines.
% use \thanks{} to gain access to the first footnote area
% a separate \thanks must be used for each paragraph as LaTeX2e's \thanks
% was not built to handle multiple paragraphs
%

\author{% <-this % stops a space
William~Lemaire, Esmaeil Ranjbar Koleibi, Maher Benhouria, Konin Koua, Jérémy Ménard, Keven Gagnon, Charles Quesnel, Louis-Philippe Gauthier, Takwa Omrani, Montassar Dridi, Mahdi Majdoub, Marwan Besrour, Sébastien Roy, Réjean Fontaine% <-this % stops a space
\thanks{This work was supported in part by the Natural Sciences and Engineering Research Council of Canada (ALLRP 557252 - 20) in part by the Regroupement Stratégique en Microsystèmes du Québec and in part by CMC Microsystems.}% <-this % stops a space
\thanks{The authors are with the Interdisciplinary Institute for Technological Innovation (3IT), Université de Sherbrooke, 3000 Université Blvd., Sherbrooke (Québec) J1K 0A5, Canada (e-mail: william.lemaire@usherbrooke.ca)}% <-this % stops a space
}

\maketitle

% For peer review papers, you can put extra information on the cover
% page as needed:
% \ifCLASSOPTIONpeerreview
% \begin{center} \bfseries EDICS Category: 3-BBND \end{center}
% \fi
%
% For peerreview papers, this IEEEtran command inserts a page break and
% creates the second title. It will be ignored for other modes.
\IEEEpeerreviewmaketitle

%DIF PREAMBLE EXTENSION ADDED BY LATEXDIFF
%DIF UNDERLINE PREAMBLE %DIF PREAMBLE
\providecommand{\DIFadd}[1]{{\protect\color{blue}\uwave{#1}}} %DIF PREAMBLE
\providecommand{\DIFdel}[1]{{\protect\color{red}\sout{#1}}}                      %DIF PREAMBLE
%DIF SAFE PREAMBLE %DIF PREAMBLE
\providecommand{\DIFaddbegin}{} %DIF PREAMBLE
\providecommand{\DIFaddend}{} %DIF PREAMBLE
\providecommand{\DIFdelbegin}{} %DIF PREAMBLE
\providecommand{\DIFdelend}{} %DIF PREAMBLE
%DIF FLOATSAFE PREAMBLE %DIF PREAMBLE
\providecommand{\DIFaddFL}[1]{\DIFadd{#1}} %DIF PREAMBLE
\providecommand{\DIFdelFL}[1]{\DIFdel{#1}} %DIF PREAMBLE
\providecommand{\DIFaddbeginFL}{} %DIF PREAMBLE
\providecommand{\DIFaddendFL}{} %DIF PREAMBLE
\providecommand{\DIFdelbeginFL}{} %DIF PREAMBLE
\providecommand{\DIFdelendFL}{} %DIF PREAMBLE
%DIF END PREAMBLE EXTENSION ADDED BY LATEXDIFF

%\input{diff.tex}
\IEEEpubidadjcol
\begin{abstract}
Neural interfaces advance neuroscience research and therapeutic innovations by accurately measuring neuronal activity. However, recording raw data from numerous neurons results in substantial amount of data and poses challenges for wireless transmission. While conventional neural recorders consume energy to digitize and process the full neural signal, only a fraction of this data carries essential spiking information. Leveraging on this signal sparsity, this paper introduces a neural recording integrated circuit in TSMC 28~nm CMOS. It features an event-driven ramp analog-to-digital converter, and a spike compression module based on principal component analysis. The circuit consists of 49 channels, each occupying an on-chip area of 50~$\times$~60~\textmu m\textsuperscript{2}. The circuit measures 1370~$\times$~1370~\textmu m\textsuperscript{2} and consumes 534~\textmu W. Compression testing on a synthetic dataset demonstrated an 8.8-fold reduction compared to raw spikes and a 328-fold reduction relative to the raw signal. This compression approach maintained a spike sorting accuracy of 74.9\%, compared to the 79.5\% accuracy obtained with the raw signal. The paper details the architecture and performance outcomes of the neural recording circuit and its compression module.
\end{abstract}


\section{Introduction}

Modern neural recording systems are instrumental for in vivo neuroscience experiments, advancing treatments for neurological disorders such as vision restoration for visually impaired patients~\cite{bloch_advances_2019}, prosthesis control for spinal cord injury patients~\cite{ajiboye_restoration_2017}, and seizure prediction for epilepsy patients~\cite{kuhlmann_seizure_2018, maturana_critical_2020}.

These systems must fulfill strict requirements regarding electrode density, power consumption, and data transmission. Although dense electrode arrays are crucial for simultaneously capturing individual neuron activity and obtaining meaningful network-level data, raw neural recordings generate a substantial amount of data. This complicates wireless transmission for large arrays designed for chronic use because of the increased power consumption.

Despite many benchtop neural recording systems supporting large-scale recordings, further reduction in both power consumption and bandwidth is necessary for achieving high channel count in implantable devices while avoiding heat-induced tissue damage~\cite{wolf_thermal_2008}. Compressing spike signals before radio transmission is one promising approach, leveraging neural signal sparsity, as only a small fraction contains valuable spike information. Although neural spike recordings can be compressed to binary threshold crossings for specific cases~\cite{trautmann_accurate_2019}, other applications necessitate waveform shapes for cell type differentiation and localization using algorithms such as spike sorting~\cite{yger_spike_2018, chung_fully_2017}. Thus, on-chip spike compression becomes essential to collect the spiking shape for these applications while minimizing power consumption and data rate.

\begin{figure}[!ht]
\includegraphics[width=0.5\textwidth]{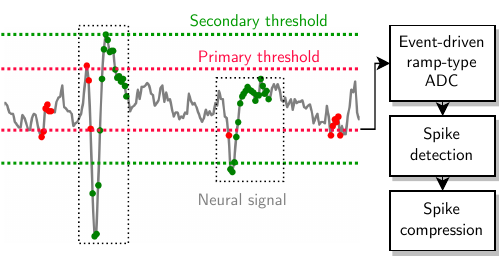}
\centering
\caption[Proposed event-driven digitization and compression scheme]{Proposed event-driven digitization and compression scheme: A ramp analog-to-digital converter begins digitizing upon crossing a primary threshold (red dots) to identify potential spike waveform onsets. If a secondary threshold is crossed within N samples, the system proceeds to capture the entire spike duration (green dots); otherwise, it resets. Following spike detection, a spike compression module applies the principal component analysis.} 
\label{fig:overview}
\end{figure}

\begin{figure*}[!ht]
\centering
\includegraphics[width=0.8\textwidth]{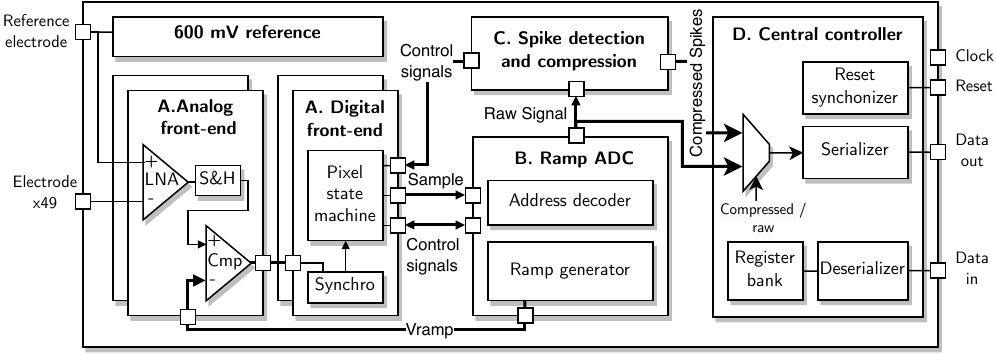}
\caption[Block diagram of the ASIC]{Block diagram of the ASIC. Each of the 49 pixels includes a front-end circuit (A.) comprising a DC-coupled LNA, a sample-and-hold circuit and a comparator. A ramp analog-to-digital converter (B.) generates a sawtooth waveform distributed to every pixel in the array. When the ramp crosses the amplified electrode signal, the comparator triggers. The readout circuit detects that transition, and outputs the digitized value and the electrode address to a spike detection and compression circuit (C.). A central controller (D.) manages the communication and configuration through a register bank. }
\label{fig:implant_overview}
\centering
\end{figure*}

On-chip compression systems can be classified into three categories. The first category involves on-chip spike sorting systems, which potentially deliver the highest compression ratio by identifying and transmiting a unique spike ID corresponding to a specific neuron. Examples of such systems include the Unsupervised Geometry-Aware OSort Clustering~\cite{chen_online-spike-sorting_2023}, discrete derivative combined with fuzzy c-means clustering~\cite{karkare_130-muw_2011} and integral transform combined with k-means clustering~\cite{do_area-efficient_2019}. Despite its efficiency, this approach is complex and generally demands extensive silicon area and memory resources. The second category encompasses systems that detect and transmit spikes without further processing~\cite{biederman_478_2015, kim_sub-wch_2019, muratore_data-compressive_2019}. While these systems are less complex and suitable for low-power, small-area implementations, they result in an inferior compression ratio. The third category integrates a compression algorithm to process detected spikes, striking a balance between compression efficiency and system complexity. Techniques employed in these systems include the discrete cosine transform ~\cite{hosseini-nejad_128-channel_2015}, discrete derivative~\cite{gibson_technology-aware_2010}, Haar discrete wavelet transform~\cite{kamboh_area-power_2007}, principal component analysis (PCA)~\cite{thorbergsson_strategies_2014}, or autoencoders~\cite{thies_compact_2019}.

These compression schemes generally involve digitizing the signal with an analog-to-digital converter (ADC), buffering it to memory, and performing spike detection and compression. To ensure the complete spike is captured and compressed, the signal is buffered before spike detection, either in the analog or digital domain.

Architectures that employ analog-domain buffering and spike detection provide notable power efficiency~\cite{kim_sub-wch_2019}. In these systems, only spike samples undergo digitization and processing, resulting in minimal digital circuit power consumption. However, the trade-off is the substantial silicon area required for the distributed analog memory. On the other hand, architectures that opt for digital-domain buffering benefit from the compact nature of digital memory~\cite{karkare_130-muw_2011}. Yet, this approach comes with a caveat: the entire signal must be digitized, buffered and processed, leading to increased power consumption.

Another approach based on a wired-or readout architecture avoids the buffer area and power penalty. It statistically discards noise prior to digitization by eliminating simultaneous samples with identical signal amplitudes. Notably, such events are more likely to occur within noise samples than spike samples, thereby providing intrinsic spike detection without the need to buffer the signal~\cite{muratore_data-compressive_2019}. Although this event-driven concept reduces power by digitizing only potentially useful spike samples, a better compression ratio potentially could be achieved by combining this idea with a digital compression algorithm, which presents a challenge because the readout circuit randomly discards samples within the spikes.

This calls for a system capable of digitizing and compressing only relevant samples, bypassing the need for a per-channel analog or digital sample buffering and reducing the number of samples that need to be digitized and processed. To leverage this opportunity, this paper introduces an event-driven ramp ADC paired with a PCA spike compression module within a 49-channel recording application-specific integrated circuit (ASIC), as illustrated in Fig. \ref{fig:overview}. The subsequent sections detail the ASIC architecture (Section II), describe the materials and methods (Section III), present the test results (Section IV) and discuss the results (Section V).

\section{ASIC Architecture}

The neural recording ASIC (Fig.~\ref{fig:implant_overview}) comprises A) 49 amplification front-ends, B) a ramp analog-to-digital converter (ADC), C) a central controller and D) a spike compression module. 

\subsection{Analog Front-End}
\begin{figure}[!ht]
\centering
\includegraphics[width=0.5\textwidth]{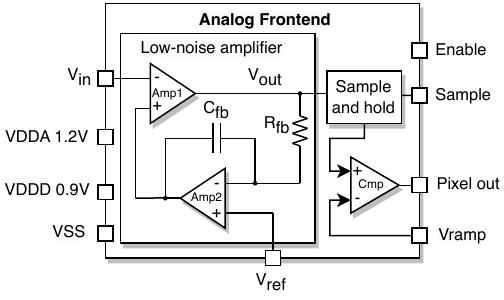}
\caption[DC-coupled low-noise amplifier (LNA) architecture]{DC-coupled low-noise amplifier (LNA) architecture. The input electrode signal (Vin) is fed to Amp1. Amp2 provides DC offset rejection by implementing a low-pass filter formed by the pseudo-resistor $R_{fb}$ and capacitor $C_{fb}$ in the feedback path of Amp1. A comparator (Cmp), coupled with a ramp generator, digitizes the signal buffered by the sample and hold circuit. The circuit uses a Cfb capacitor of 1 pF and a  sample and hold capacitor of 235 fF. }
\label{fig:pixel_diagram}
\end{figure}

The analog front-end needs to amplify neural signals while canceling the DC electrochemical voltage difference between recording and reference electrodes. This can be achieved with AC-coupled amplifiers based on capacitor feedback networks~\cite{hashemi_noshahr_multi-channel_2020}. However, this architecture necessitates a large capacitor area, hindering integration in a high-density recording array. Alternatively, DC-coupled architectures enable input offset rejection with a smaller capacitor area~\cite{gosselin_low-power_2007}. To attain the required high-pass filtering, an active low-pass filter is placed in the main amplifier's feedback loop. Due to the Miller effect, the needed capacitor is significantly smaller than in the AC-coupled configuration. Fig.~\ref{fig:pixel_diagram} provides a high-level overview of the amplifier circuit. A separate publication details the front-end architecture~\cite{koleibi_low-power_2022}.

\subsection{Ramp ADC}

The neural signal sparsity presents an opportunity for power savings within the analog-to-digital conversion process. Since a typical successive-approximation register (SAR) ADC continuously operates regardless of spike presence, sampling exclusively during spike events can reduce power. Additionally, the high number of channels suggests potential for power and area improvement through hardware sharing.

\begin{figure}[!ht]
\centering
\includegraphics[width=0.5\textwidth]{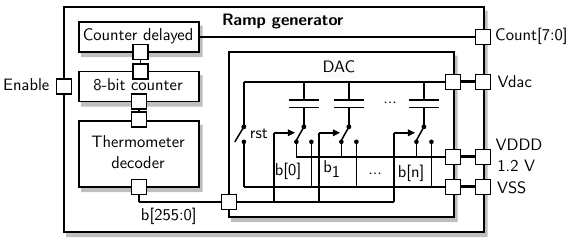}
\caption[Ramp generator circuit]{Ramp generator circuit based on a  unary-weighted split-capacitor array. An 8-bit counter is converted by a thermometer decoder to a 256-bit code and fed to a unary-weighted split-capacitor array. }
\label{fig:ramp_generator}
\end{figure}

Considering these constraints, the proposed circuit includes an event-driven ramp ADC. A digital-to-analog converter (DAC) generates a ramp shared by all channels. The ramp generator employs a unary-weighted split-capacitor array~(Fig.~\ref{fig:ramp_generator}) for improved linearity, albeit at the expense of increased silicon area when compared to a binary-weighted array. As there is only a single ramp shared by all pixels, the larger area is acceptable. The ramp output is routed to 49 comparators, each situated in a separate channel. When the ramp voltage crosses LNA output in a specific channel, the readout circuit latches the value of the global 8-bit counter controlling the ramp. 

In contrast to traditional ramp ADCs, the proposed scheme integrates a dual-threshold mechanism to avoid digitizing non-spike samples. The ADC only initiates digitization if the signal amplitude is above the first threshold. Following this, if the signal crosses a second threshold within a predetermined number of samples, the spike is retained; otherwise, it is discarded. This scheme avoids unnecessarily
spending power digitizing samples below the noise threshold.

\begin{figure}[!ht]
\centering
\hspace*{-0.30cm}\includegraphics[width=0.51\textwidth]{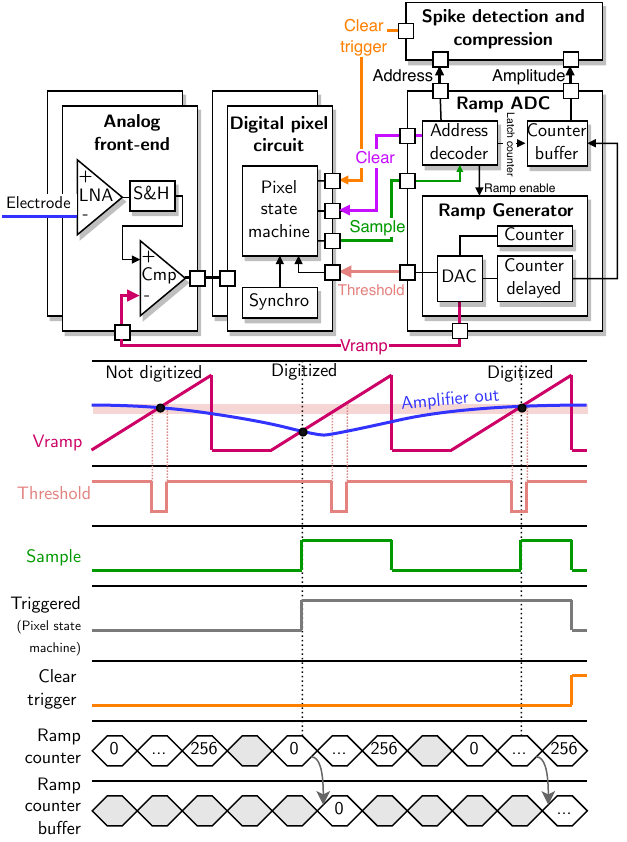}
\caption[Event-driven digitization scheme]{Event-driven digitization scheme: A first event below thresholds is not digitized. A second event, above thresholds, is digitized and triggers the pixel. A third event below threshold is digitized due to the triggered pixel. Lastly, the spike detector circuit clears the trigger after a configurable time window (clear trigger signal). }
\label{fig:sequence_diagram}
\end{figure}

Fig.~\ref{fig:sequence_diagram} presents the operating principle of the event-driven ADC. In event-driven mode, the digital front-end only enables the ADC after the signal crosses a configurable amplitude threshold (\textit{Threshold} signal in Fig.~\ref{fig:sequence_diagram}). Subsequently, the readout circuit clears the trigger if no spike is detected within a configurable time window. Otherwise, if the signal reaches a second amplitude threshold within this interval, the spike detector records a full spike by waiting a presettable time before clearing the trigger. 

\begin{figure}[!ht]
\centering
\includegraphics[width=0.35\textwidth]{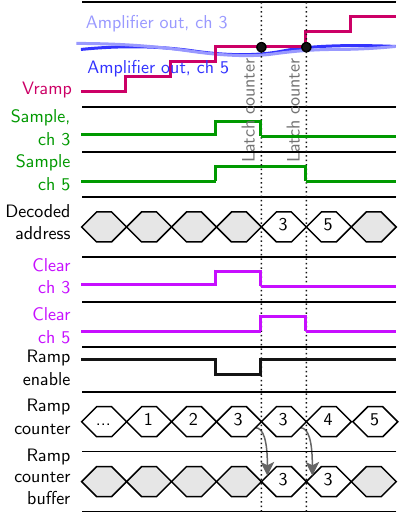}
\caption[Collision resolution with the shared ramp ADC]{Collision resolution with the shared ramp ADC when multiple channels present a similar voltage level, necessitating the digitization of several values within a single ramp step.}
\label{fig:resolving_collisions}
\end{figure}

When multiple electrodes exhibit similar voltages, a collision occurs due to the ramp ADC's characteristic of resolving only a single sample per clock cycle. The mechanism for resolving collisions is detailed in Fig.~\ref{fig:resolving_collisions}. When detecting a collision, the \textit{Address decoder} deasserts the \textit{Ramp enable} signal to pause it. Subsequently, the ADC digitizes and clears the pending channels (\textit{Clear ch} signal) in sequence, beginning with the channel with the lowest address, before reactivating the ramp.

\subsection{Spike Compression}
Neural recording circuits typically operate at sample rates between 20 kHz and 30 kHz, with resolutions ranging from 8 to 16 bits~\cite{ballini_1024-channel_2014, zhang_closed-loop_2015, delgado-restituto_system-level_2017}. This results in a data bandwidth of 160 kbit/s to 480 kbit/s per electrode. The challenge arises when attempting to transmit data from multiple channels using a low-power radio transmitter. To address this issue, the paper introduces a combination of an event-driven ADC with a spike detection and compression module.

The spike detection module incorporates a dual threshold detector, which works in conjunction with the ADC. Activation of the ADC for digitization occurs upon reaching the first threshold. If the signal surpasses a second threshold within a designated number of samples, the compression module retains the spike. Otherwise, it discards the spike. After detecting the spike, the system compresses it using principal component analysis.

\begin{figure}[!ht]
\includegraphics[width=0.50\textwidth]{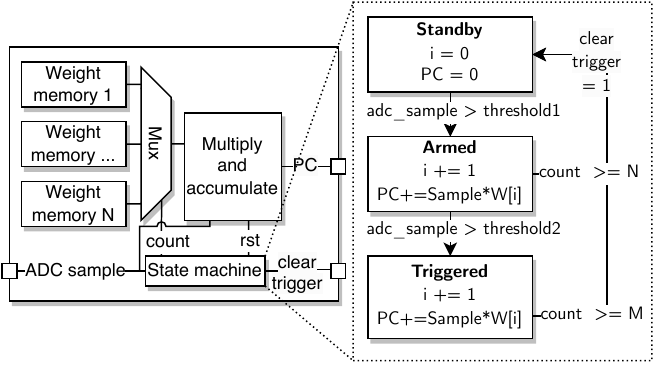}
\caption[block diagram of the spike detection and compression unit]{Simplified block diagram of the spike detection and compression unit.}
\label{fig:compression_algorithm}
\end{figure}
The compression algorithm employs a feature extraction module operating on the detected spikes. PCA has been selected for its proven efficiency in spike sorting feature extraction and its relative simplicity for hardware integration~\cite{chen_efficient_2015, tung-chien_chen_vlsi_2008}. 

In contrast to typical spike sorting techniques that calculate PCA separately for each electrode, the proposed compression technique computes PCA coefficients on a matrix that concatenates detected spike waveforms across all channels in its columns. This removes the requirement for per-channel memory and allows for the compression of a broad range of spike shapes with a single set of coefficients. A related study tested a fixed weight set, derived from a vast collection of experimental spike waveforms, as a universal compression basis. The study observed no significant degradation of the spike sorting performance when compared to an optimal dataset-specific compression basis~\cite{thorbergsson_strategies_2014}. These findings support the approach undertaken in this paper comprising a single set of weights for all electrodes.

%By combining multiple datasets for calculating PCA coefficients, the method can potentially achieve better robustness against dynamic changes in neural recordings compared to a per-channel PCA. 
The implemented PCA compression scheme is as follows:

\begin{equation}
P = X W
\end{equation}

\noindent where $P$ is the resulting principal components, $X$ a row vector containing samples, and $W$ is the matrix of principal component coefficients. The compression subsystem quantizes the four principal components to 6 bits each, resulting in a total of 24 bits. The decompression with the inverse PCA is as follows:

\begin{equation}
\mathrm{Decompressed\_spikes} = P W^{T}
\end{equation}

Calculating the principal component coefficients involves computing the covariance matrix of the neural spikes dataset, and performing eigen-decomposition. The matrix $W$ is formed from the $k$ eigenvectors with the highest eigenvalues. Reducing $k$ increases the compression ratio by using fewer principal components to represent the data, but also loses more information, thereby decreasing the reconstruction quality.

Fig.~\ref{fig:compression_algorithm} shows the compression algorithm implementation in digital hardware. When the signal crosses the primary threshold, the module starts receiving ADC samples. A multiply-and-accumulate block sequentially computes each sample's contribution to the principal component's partial sum. Over a span of N recorded samples, if the signal meets or exceeds the second threshold, the state machine proceeds to process an additional M samples before resetting the trigger. Otherwise, it clears the trigger and the partial sum immediately. In the calibration phase, a grid search optimizes spike sorting accuracy by tuning the pretrigger sample quantity (N) and the post-trigger sample quantity (M). The experiments conducted use a pretrigger sample quantity (N) of 3 and a post-trigger sample quantity (M) of 19. 

In scenarios with simultaneous spiking events, the compression hardware processes interleaved samples from multiple channels in real time. For each incoming sample, it reads three elements from the channel memory: the channel's state (2 bits), the sample index $i$ (5 bits), and the running sum of each PCA component (4~$\times$~11 bits). Subsequently, it retrieves the PCA coefficients corresponding to the current sample index $i$ from the PCA memory (4 components~$\times$~22~samples/spike~$\times$~9 bits). The hardware then multiplies each sample by its respective PCA coefficients and adds it to the PCA running sum. The hardware determines the next action based on the channel status: if the spike has ended, it forwards the compressed spike (PCA running sum) for serialization; if the spike is ongoing, it updates the channel memory with the new channel state, spike index, and PCA running sum. This circuit processes a new sample every 16 MHz clock cycle. Despite being paired with only 49 electrodes in the current design, the compression circuit could handle data from up to 800 channels with simultaneous spiking activity. The memory requirements for this design amount to 51 bits per channel and 792 bits for PCA coefficients, leading to a total memory requirement of 0.4 kB. The ASIC supports both raw and compressed recording modes for calibration of detection thresholds and PCA coefficients.

\subsection{Central Controller}

The ASIC central controller handles bidirectional communication, including configuration register operations and transmission of neural data. 

The configuration registers include individual pixel enable, primary and secondary threshold values, sampling period, compression weights, and pretrigger sample quantity (N).

For implantable communication wires, the inbound communication uses AC coupling and Manchester encoding to eliminate DC levels. The signaling occurs at 1.2~V amplitude, with a 2 Mbit/s data rate. The outbound communication signaling occurs at 1.2~V amplitude, with a 16 Mbit/s data rate. %Data packet types include register values, raw neural signals (Fig.~\ref{fig:raw_packet}) or compressed spikes (Fig.~\ref{fig:compressed_packet}). 

%\begin{figure}[!ht]
%\includegraphics[width=0.50\textwidth]{Media/raw_packet.pdf}
%\caption[Raw packet format]{Raw packet comprising 1-30 samples, each including an address and ADC value. The number of samples for a given timestamp only depends on the number of active electrodes when spike detection is disabled and scales with the neural activity when the spike detection is enabled.}
%\label{fig:raw_packet}
%\end{figure}

%\begin{figure}[!ht]
%\includegraphics[width=0.50\textwidth]{Media/compressed_packet.pdf}
%\caption[Compressed packet format]{Compressed packet comprising 1-12 spike samples includes an address and the spike shape represented by a 4-dimensional PCA vector. When the packet contains multiple spikes, the timestamp, electrode address and principal components are repeated. }
%\label{fig:compressed_packet}
%\end{figure}

\section{Materials and Methods}

The ASIC is characterized for ADC performance, compression performance, spike sorting accuracy, and power consumption. The following subsections provide detailed explanations for each measurement.

\subsection{ADC Linearity}

The ADC's linearity is measured via a code density test, identifying the differential non-linearity (DNL) and integral non-linearity (INL). The linearity test is achieved by repeatedly injecting a ramp reference signal into the ADC, and measuring the deviation of hits per ADC codes from a uniform distribution. A 14-bit AD9717 DAC with ±1.8 least significant bit INL and 0.2~\si{\milli\volt} resolution repeatedly injects a 1 Hz ramp signal from 0 to 1~\si{\volt} into the ADC. 

DNL is computed by dividing the hits per ADC bin $i$ by the average hits across all codes, highlighting deviations from a uniform distribution:

\begin{equation}
\label{eq:dnl}
\mathrm{DNL}(i) = \frac{\mathrm{Number\_of\_Hits}(i)}{\mathrm{Average\_Number\_of\_Hits}} - 1
\end{equation}

The ADC's deviation from ideal linearity is quantified by the integral non-linearity (INL), which is obtained from the cumulative sum of the differential non-linearity (DNL):
\begin{equation}
\label{eq:inl}
\mathrm{INL}(i) = \sum_{n=0}^{i} \mathrm{DNL}(n)
\end{equation}

\subsection{Compression Performance} 

Compression performance is evaluated by introducing a pre-recorded neural dataset into the electrode input via a digital-to-analog converter. The spike sorting accuracy is subsequently compared between the original and compressed datasets.

For this purpose, a dataset containing the verified timings of neuron spikes is necessary, also referred to as the ground truth. While real datasets are ideal, they typically contain very few neurons with such timing information due to the laborious procedure of recording the ground truth with intracellular recordings. Therefore, synthetic datasets are used, as they provide a higher quantity of neuron recordings paired with the ground truth data.

A synthetic dataset is generated using MEArec, a software package designed for creating synthetic extracellular neural recordings on multi-electrode arrays~\cite{buccino_mearec_2021}. Recordings are constructed by convolving pre-recorded extracellular action potential templates with random spike trains and subsequently adding noise. The characteristics of the generated dataset are presented in Table \ref{tbl:synthetic_dataset}.

\begin{table}[!ht]
\centering
\caption{Synthetic dataset characteristics}
\label{tbl:synthetic_dataset}
\begin{tabular}{ll}
\Xhline{2\arrayrulewidth}
\textbf{Cell model quantity} & 130 \\
\textbf{Templates per cell} & 100 \\
\textbf{Recording duration} & 20 s \\
\textbf{Noise level} & 10 \si{\micro\volt} \\
\textbf{Bandwidth} & 500 to 6000 Hz \\
\Xhline{2\arrayrulewidth}
\end{tabular}
\end{table}

The dataset, split into training and test sets, is introduced into the electrode input using a 16-bit AD5686BRUZ digital-to-analog converter. A resistive bridge attenuates the voltage by a factor of 1000 to match the electrode input's voltage range and reduce DAC-generated noise. During the initial calibration run, the training dataset is introduced into the electrode, and the ASIC amplifies, digitizes, and transmits the raw samples to a Xilinx 7000-series Zynq system-on-module. Compression weights are computed and sent back to the ASIC. Finally, the test dataset is injected, and the ASIC transmits back the compressed representation. 

Spike sorting is performed on the recorded raw and compressed datasets using the trisdesclous framework via SpikeInterface~\cite{buccino_spikeinterface_2020}. The spike sorting performance is evaluated based on spike sorting accuracy:

\begin{equation}
\mathrm{Accuracy} = \frac{\mathrm{TP}}{\mathrm{TP} + \mathrm{FN} + \mathrm{FP}}
\end{equation}

\noindent where TP represents the true positives (a sorted spike well matched to a ground-truth spike), FN the false negatives (an unmatched ground truth spike) and FP the false positives (an unmatched sorted spike).

\textbf{\begin{figure}[!ht]
\includegraphics[width=0.5\textwidth]{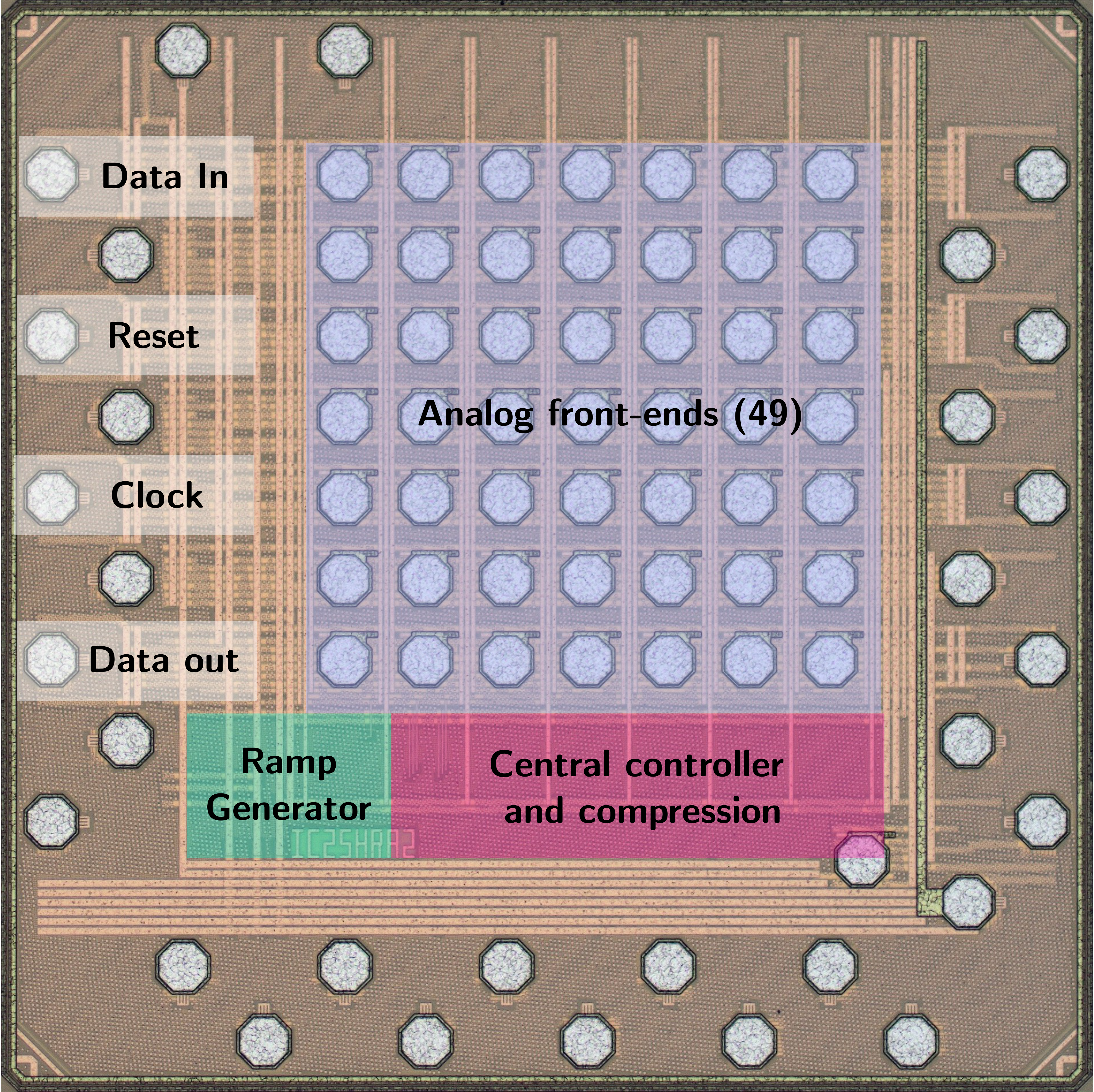}
\caption{Microphotograph of the Neural Recording ASIC.}
\label{fig:implant_layout}
\end{figure}}

\subsection{Power Consumption}

The power consumption is measured for both analog modules (LNA, comparator, bias generator, ramp generator, reference electrode buffer and bandgap) and digital modules (digital controller). For the analog modules, the power consumption is measured with a DC simulation in Cadence Spectre APS. For the digital circuit, the power consumption (switching, internal, leakage) is measured with a combination of Cadence Incisive (for running the simulation) and Cadence Innovus (for extracting post-routing power consumption from the simulation). Finally, the analog front-end and digital circuit power consumption are measured on the fabricated ASIC using a TI INA226 power monitor.

The analog supply powers test circuits not intended for a final implantable chips (analog buffers). Thus, the power consumption of these circuits is simulated and subtracted from the total analog power, both measured and simulated.

\section{Results}

The layout of the fabricated ASIC, measuring 1370~$\times$~1370~\si{\micro\meter\squared}, is presented in Fig.~\ref{fig:implant_layout}. The results are presented for ADC linearity, compression performance, and power consumption.

\subsection{ADC Linearity}

\begin{figure}[!ht]
\includegraphics[width=0.50\textwidth]{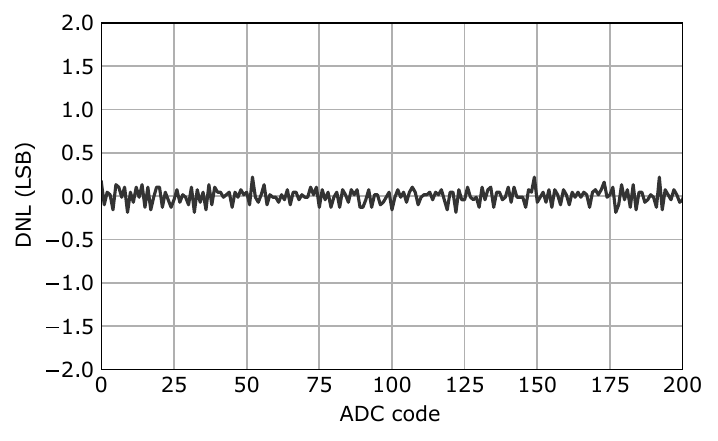}
\caption{Differential non-linearity of the ADC. }
\label{fig:dnl}
\end{figure}

Figure \ref{fig:dnl} presents the differential non-linearity of the ADC. Since the ADC doesn't operate through the full rail-to-rail voltages, it yields fewer valid ADC codes compared to the full 8-bit range. Codes outside the operating range are not represented.

\begin{figure}[!ht]
\includegraphics[width=0.50\textwidth]{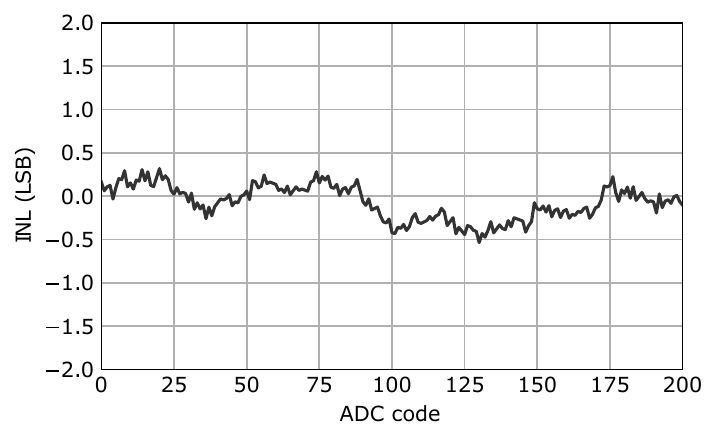}
\caption{Integral non-linearity of the ADC. }
\label{fig:inl}
\end{figure}

Figure \ref{fig:inl} presents the integral non-linearity. The INL has a deviation below 0.5 bins, owing mainly to the ramp ADC architecture.

\subsection{Compression Performance}

\begin{figure}[!ht]
\includegraphics[width=0.50\textwidth]{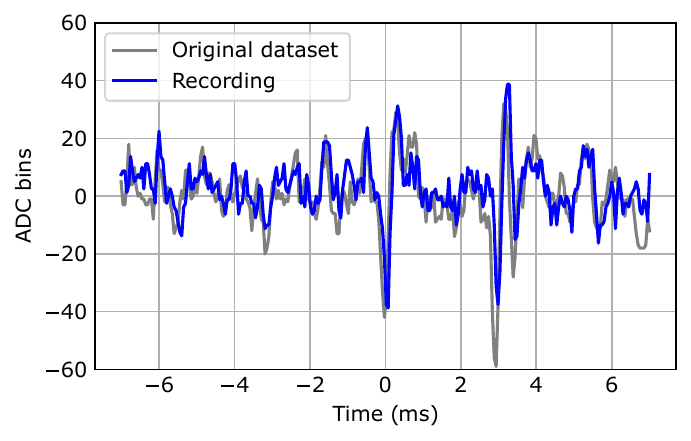}
\caption[Spike sorting on the raw and compressed datasets]{Comparison of the recorded and digitized signal (recording) to the original synthetic dataset produced with MEARec.}
\label{fig:raw_recording}
\end{figure}

\begin{figure}[!ht]
\includegraphics[width=0.50\textwidth]{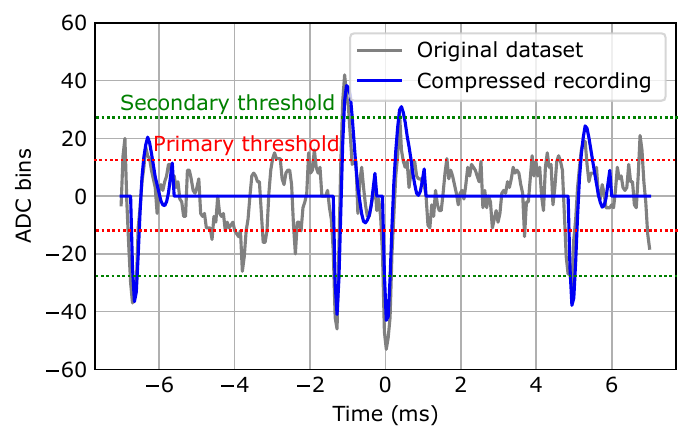}
\caption[Signal compression example]{Signal compression on a synthesized MEARec dataset compared to the original dataset. The dataset is injected with a digital-to-analog converter and passes through the amplification front-end, ramp ADC and spike compression.}
\label{fig:compressed_recording}
\end{figure}

Fig.~\ref{fig:raw_recording} presents the raw recording of a randomly selected spike signal from the MEARec dataset. Fig.~\ref{fig:compressed_recording} presents examples of compressed and reconstructed spikes.

\begin{figure*}[!ht]
\includegraphics[width=1.0\textwidth]{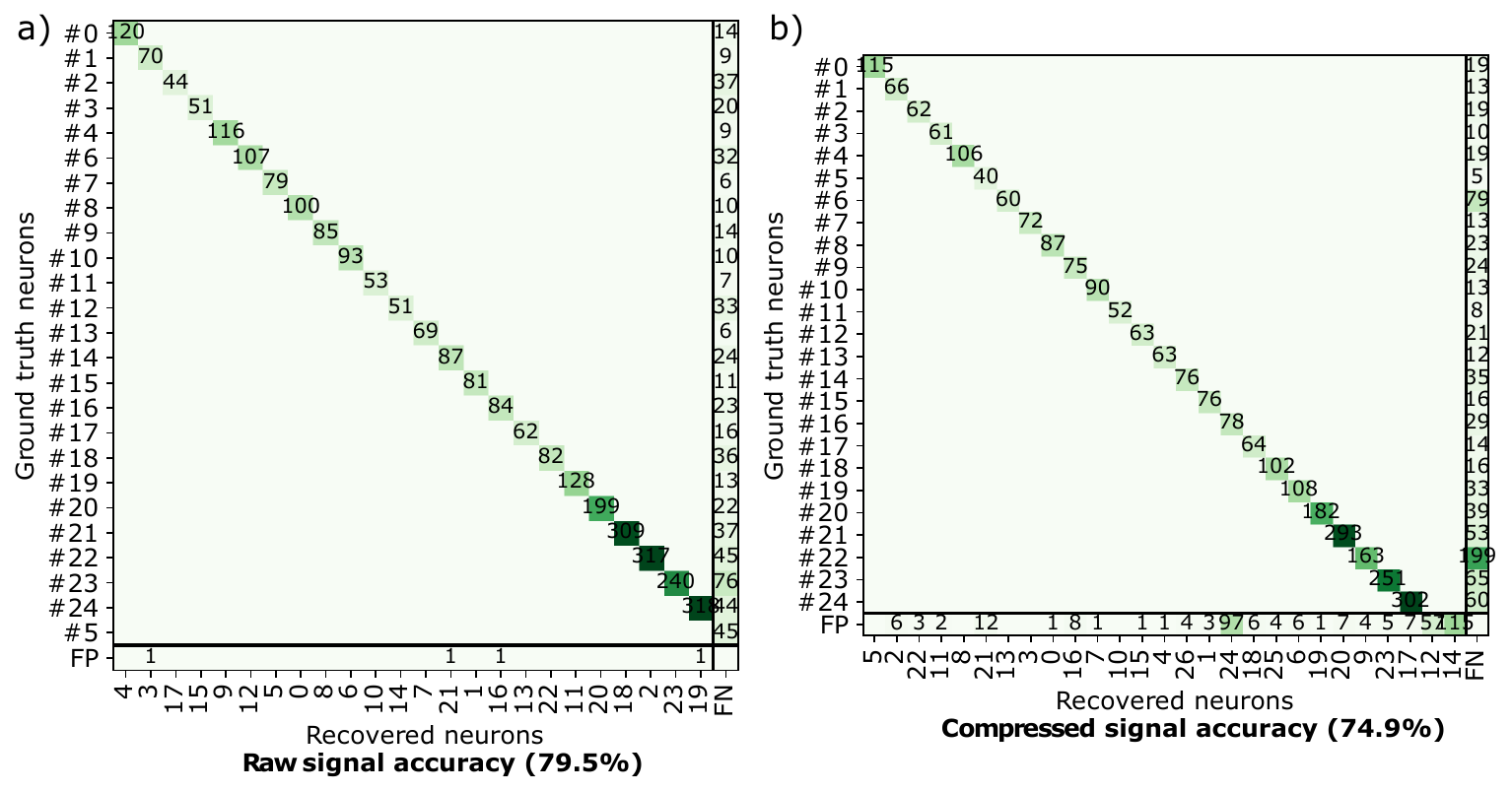}
\caption[Confusion matrices for spike sorting results]{Confusion matrices for spike sorting results of (a)  raw signals, and (b) compressed signals. Rows denote IDs of ground truth neurons and columns represent IDs of neurons identified by the spike sorting algorithm. Cell values indicate matching spike counts within a defined time window. The right-hand column shows unobserved firings of ground truth neurons (false negatives), and the bottom row reports unmatched firings from recovered neurons.}
\label{fig:confusion_matrix}
\end{figure*}

The compression system's effectiveness at preserving critical information is evaluated through spike sorting and by comparing its outcomes with ground truth data. These results are encapsulated in a confusion matrix, as depicted in Fig.~\ref{fig:confusion_matrix}, which visually represents the performance of a classification algorithm by displaying the number of correct and incorrect predictions. In this context, each row of the matrix denotes a neuron from the ground truth dataset, while each column corresponds to a neuron identified by the spike sorting algorithm. The raw signal achieves a sorting accuracy of 79.5\%, in contrast to 74.9\% for the compressed signal. Table \ref{tbl:compresson_performance} presents the compression ratios and spike sorting accuracies for raw signal, detected spikes and compressed spikes.

\begin{table}[!ht]
\centering
\caption{Compression performance}
\label{tbl:compresson_performance}
\begin{tabular}{lll}
\Xhline{2\arrayrulewidth}
\textbf{Dataset} &  \textbf{Compression ratio} &  \textbf{Spike sorting accuracy} \\
\hline
MEARec Raw & 1 & 79.5\% \\
MEARec Detected & 37 & 79.5\% \\
MEARec Compressed & 328 & 74.9\% \\
\Xhline{2\arrayrulewidth}
\end{tabular}
%\end{adjustbox}
\end{table}

\subsection{ASIC Power Consumption}

Table~\ref{tbl:power_consumption} presents the power consumption of the various modules of the chip, obtained via simulations and real-world measurements as detailed in subsection III.A.

\begin{table}[!ht]
    \centering
    \caption{Power consumption of digital and analog circuits}
    \label{tbl:power_consumption}
    \begin{tabular}{lll}
        \toprule
        \textbf{Module} &  \textbf{Simulated (\si{\micro\watt})} & \textbf{Measured (\si{\micro\watt})} \\
        \midrule
        \textit{Digital:} & & \\
        \quad Controller + compression &  69 (internal) + &   \\
        \quad  &  37 (switching) + &   \\
        \quad &   22 (leakage) &  \\
        \textbf{Digital total} & \textbf{128} & \textbf{121} \\
        \addlinespace
        \textit{Analog:} & & \\
        \quad LNA + comparator ($\times$49) &  153 (LNA) + &  \\
        \quad & 130 (comp) &  \\
        %\quad Bias generator &  77 &  \\
        %\quad Reference electrode buffer &  22 &  \\
        %\quad Bandgap &  10  &   \\
        \quad Bias generator &  80 &  \\
        \quad Reference electrode buffer &  100 &  \\
        \quad Bandgap &  13  &   \\
        \quad Ramp generator &  52 &  \\
        \quad Test circuits (excluded) &  344  &   \\
        \textbf{Analog total (w/o test circuits)} & \textbf{528} & \textbf{413} \\
        \midrule
        \textbf{Total} &  \textbf{656} & \textbf{534} \\
        \bottomrule
    \end{tabular}
    \footnotetext{The power consumption of test circuits is listed separately and is not included in the Analog total or the overall Total.}

\end{table}

\begin{table*}[!ht]
\caption{Comparison of neural recording ASICs with embedded spike compression}
\begin{adjustbox}{width=1\textwidth}
\begin{tabular}{lllllll}
\Xhline{2\arrayrulewidth}
 &	\textbf{This work} &	\textbf{Zhang, 2016}~\cite{zhang_closed-loop_2015} & \textbf{Biederman, 2015}~\cite{biederman_478_2015} & 	\textbf{Kim, 2019}~\cite{kim_sub-wch_2019} & \textbf{Hao, 2021} \cite{hao_108_2021} &  \textbf{Zeinolabedin, 2022} \cite{zeinolabedin_16-channel_2022}\\
\hline
Technology & 28 nm &	180 nm &	65 nm &	180 nm & 180 nm & 22 nm FDSOI\\
Number of amplifiers &	49 & 4 &	64 &	16 & 4 & 16\\
Supply voltage analog (V) / digital (V) &	1.2 / 0.9 &	0.5 / N/A &	1.0 / 0.8 &	0.5 / N/A & 1.5 &  0.8  \\
Signal bandwidth (kHz) & 0.5 to 9.2 &	0.5 to 6 & N/A	 &	6.8 & 0.4 to 5 & 0.8 to 9.8\\
Input-referred noise (\textmu V RMS) &	15.8 &	3.1 &	7.5 &	5.4 & 3.3 & 5\\
Gain (dB) &	54.6 &	40 to 58 &	48 &	59 & 44 to 73 & 28 to 52\\
ADC type & Event-driven ramp &  Shared SAR	 & Shared SAR & Per-channel SAR & ADC-less & Shared SAR  \\
ADC resolution (bits) & 8 &	10 & 10 & 8 & N/A & 9 \\
Sampling rate (kHz) & 20 &31.25 & 20 &	31.25 & N/A & 25 \\
Compression system & PCA & Compressed sensing & Spike detection &	Spike detection only & FSDA & Configurable \\
Digital power (\textmu W) &	121 &	N/A & N/A &	N/A & N/A & N/A \\
Total power (\textmu W) &	534 & 965 (with UWB)&	N/A &	14.1 & 10.8 & N/A \\
Total area (mm\textsuperscript{2}) & 1.4~$\times$~1.4 & 5~$\times$~5 & 2.15~$\times$~2.225 & 2.35~$\times$~2.5 & 2.79 & N/A \\
\hline
Compression ratio @ mean spike rate & \textbf{328 @ 20 Hz} &  10.6 @ N/A & 8.3 @ 50 Hz & 9 @ N/A & N/A / N/A &  100 @ 100 Hz \\
%Digital area (um\textsuperscript{2}) & 64,000 &  &  &  &  & \\
Digital area per channel (\textmu m\textsuperscript{2}) & \textbf{1300}  & 110,000  & 10,500 &  18,000 (detection only) & 1,023,000 & 14,000 \\
%Digital power per channel (\textmu W) & 2.46  & \textbf{0.83}   &   &  & 4.07 & 2.79 \\

Front-end power per channel (\textmu W) &	2.5 &	15 &	3.3 &	\textbf{0.88} & 6.1 & 1.52 \\
Front-end area per channel (\textmu m\textsuperscript{2}) &	\textbf{3,000 (50$\times$60)}  &	550,000 & 25,800 &	160,000 (200$\times$800) & 357,000 & 38,000\\
\Xhline{2\arrayrulewidth}
\end{tabular}
\end{adjustbox}
\label{tbl:comparison}
\end{table*}

\section{Discussion}

\subsection{Event-driven Ramp ADC: Benefits and Limitations}

The event-driven ramp ADC combines most of the strengths and mitigates some of the weaknesses of traditional spike detection and buffering methods involving analog or digital memory.

Analog buffering and spike detection efficiently digitize full spike waveforms, including pre-trigger samples, avoiding the energy cost of digitizing unnecessary inter-spike samples. Their requirement for considerable per-channel area for the analog buffer capacitors, however, renders them difficult to apply to large, dense electrode arrays.

Conversely, the digital buffer approach calls for the digitization, storage and processing of all samples, resulting in elevated power consumption. Given that spike samples often constitute a very small fraction of the dataset, this method inefficiently expends additional energy for storing results in SRAM memory or equivalent. 

The event-driven ramp ADC combines the advantages of both methods, enabling digitization of only spike signals with a compact in-pixel circuit and minimal memory use. However, an area for improvement lies in the accurate recording of spike onset, which may contain crucial features beneath the first threshold. This limitation pertains to applications that identify low-amplitude spike features by averaging multiple waveforms from the same neuron~\cite{tandon_automatic_2021}.

\subsection{CMOS Technology Choice}
Very few 28~nm CMOS neural recording circuits are reported in the literature, which contrasts with numerous 65-nm~\cite{wu_streaming_2017, cuevas-lopez_low-power_2022}, 130-nm~\cite{brenna_64-channel_2016, cuevas-lopez_low-power_2022} and 180-nm CMOS efforts~\cite{kim_sub-wch_2019}. The use of smaller technology nodes complicates the design, but offers opportunities for area and power reduction. However, higher leakage is a drawback of the planar TSMC 28 nm HPC technology. This characteristic becomes more pronounced for low-frequency digital circuits such as the one presented, but can be partly mitigated using high-voltage threshold (HVT) digital cells. 

\subsection{Experiment Limitations}

The experiments for evaluating spike sorting methods necessitate ground-truth datasets. These datasets pair extracellular and patch-clamp recordings to provide ground-truth spike times. However, the pairing method is typically limited to a few cells because of the manual work required. As a solution, synthetic recordings with a large number of cells can be produced by convolving real spike templates with modeled noise sources.

Although synthetic datasets yield a substantial number of neurons with ground-truth data, they fall short in perfectly modeling neuron bursting behaviour, noise sources, and motion artifacts. Despite these limitations, synthetic datasets effectively demonstrate the system's capability to compress spikes while preserving essential information for spike sorting.

Additionally, the experiments are realized by injecting a synthetic dataset with a DAC. The DAC generates a signal with low output impedance, contrasting the high input impedance of electrodes. Further validation tests are necessary to ensure minimal impact on the results.

The compression ratio between raw spikes and raw signal exhibits a strong dependence on the neural activity present in the dataset. In this study, the synthetic dataset was configured to reproduce a mean spike rate of 20 Hz. The compression ratio between compressed spikes and raw spikes is independent of the dataset and is not impacted by the average spike rate.

\subsection{System Scalability}
The proposed system is a step towards addressing the future demands of neural interfaces: denser electrodes, lower power consumption, and decreased data rates compatible with wireless transmission. Event-driven digitization stands as a viable solution for scalability due to the characteristics of neural signals, where spiking information represents only a small fraction of samples~\cite{muratore_data-compressive_2019}. This method reduces power consumption by minimizing the requirements for sample digitization and spike detection.

The proposed event-driven ADC scheme enhances scalability by removing memory buffers. A large 10,000-electrode array and a 10-bit ADC with 10 pre-threshold samples would require 1 Mbit of memory, increasing area and power usage. The dual threshold spike detection method allows to partially detect the spike onset without any signal buffering.

Although other published systems also utilize event-driven digitization, they often lack additional signal compression~\cite{kim_sub-wch_2019, muratore_data-compressive_2019}, which is crucial for wireless systems requiring full spike waveform.

Additionally, the proposed system's scalability potential rests in part on the concentration of most analog circuitry within the pixel, leaving only a noise-insensitive digital comparator output. Accommodating additional pixels and wires mostly requires scaling the in-pixel comparators and the ramp ADC buffer drive strengths. However, routing congestion could arise from the one-to-one wiring between pixel output and digital circuitry. Although the nine metal layers provided by the 28 nm technology provide high routing density, this scheme can potentially cause routing congestion as the pixel count increases. An integrated row/column decoding scheme for the pixel comparator output, as exemplified by Muratore et al. ~\cite{muratore_data-compressive_2019}, might offer a solution.

The ramp ADC in the proposed system also demonstrates potential for scalability. Each digitized sample necessitates a single clock cycle per sampling period. Within a 50 \si{\micro\second} (20 kHz) sampling period with a 16 MHz system clock, up to around 800 samples can potentially be digitized. Clock cycles are only utilized when a spike event triggers the ADC, enabling a single ADC to manage a significantly larger number of channels compared to non-event-driven systems. Further expansion in pixel count could be achieved through additional parallel ADCs or by increasing the clock frequency.

The compression module exhibits scalability potential as well. It can process one spike sample per 16 MHz clock cycle. Given a 20 kHz sampling rate and a spike density of 1 sample out of 40~\cite{muratore_data-compressive_2019}, the compression system has the potential to theoretically handle 32,000 channels. However, due to practical limits, this figure will be lower due to the stochastic spike event distribution. Ultimately, designing and fabricating a large scale array will be necessary to evaluate the challenges involved.

\subsection{Comparison With Other Works}
Table~\ref{tbl:comparison} compares the ASIC with other published circuits. 
\subsubsection{Area and Power}
Notably, many neural circuits utilize larger-scale technology nodes such as 130~nm, 180~nm, and 350~nm. The use of 28~nm CMOS technology in this study provides a favorable balance between compression ratio and area, particularly for digital circuits. Even though some circuits exhibit lower front-end power per channel, the front-end area of the design presented here is significantly smaller than that of the other circuits in Table~\ref{tbl:comparison}.  Additionally, the event-driven digitization, detection, and compression approach presented in this paper stands out by eliminating signal buffering, a common feature in other designs. This method digitizes only relevant spikes in real time using a dual-threshold detector, leading to benefits in power and area. This results in a compact 64,000~\si{\micro\meter\squared} digital area. In comparison, designs comprising signal buffering generally consume significant power and require a large silicon area. For instance, one referenced design requires 64~\% of its total power for memory and occupies an area of 504,100~\si{\micro\meter\squared} in 22~nm CMOS \cite{chen_online-spike-sorting_2023}, while another uses 51~\% of its 3,060,000~\si{\micro\meter\squared} digital area and 71~\% of its power for memory. Similarly, a third design achieves very low power on-chip spike sorting, but still requires a significant area of 414,000, and 64~\% of the energy is spent buffering and detecting spikes~\si{\micro\meter\squared}~\cite{do_area-efficient_2019}.

\subsubsection{Spike Sorting Accuracy}
While some papers report superior outcomes for spike sorting accuracy~\cite{chen_online-spike-sorting_2023, do_area-efficient_2019, zeinolabedin_16-channel_2022}, this metric highly depends on the selected spike sorting algorithm and dataset. Direct comparison across methods is not realized because the papers compared all use different datasets. This paper instead focuses on validating that compression does not significantly degrade sorting accuracy, reducing dependence on specific datasets and sorting algorithms.

\subsubsection{Compression Ratio}
The system described in this paper opts for spike compression over on-chip spike sorting to minimize power and area requirements while still providing a significant compression ratio. Alternatively, on-chip spike sorting can potentially achieve higher compression ratios by transmitting only spike IDs per channel~\cite{do_area-efficient_2019}, but at the cost of more complexity. Similarly, other complex compression schemes like convolutional autoencoders can also offer better compression ratios, but at the cost of increased silicon area requirements. For example, a convolutional autoencoder compression system uses 12,000,000~\si{\micro\meter\squared} of silicon area~\cite{seong_multi-channel_2021}. Other compression systems use spike detection only and provide much lower compression ratios such as 10.6~\cite{zhang_closed-loop_2015}, 8.3~\cite{biederman_478_2015} or 9~\cite{kim_sub-wch_2019}. This comparison underlines the trade-offs between complexity and compression efficiency in neural signal compression systems.

Furthermore, the proposed design compares to another event-driven approach utilizing a single ramp ADC \cite{muratore_data-compressive_2019}. However, this architecture provides spike detection without compression, whereas the proposed method potentially offers an 8.8 times greater compression ratio, a significant advantage for neural implants where wireless transmission limitations impact channel count.

% Spike detection only (biederman, Kim) small digital area, but limited compression ratio
% Compressed sensing on the whole signal (Zhang): Lower compression ratio than for spike-driven approaches
% Full analog spike sortier (Hao), very large area per channel
% Flexible approach, very large area per channel. 
% 
% On-chip spike sorting. 
% With SRAM fifo, large area, consumes 64% of the total power https://ieeexplore.ieee.org/stamp/stamp.jsp?tp=&arnumber=10231079 685 µW for 384 channels. 
% spike detector consumes 39%. Much larger area for digital circuit 504,100 um2 vs 64,000 um2
% On-chip spike sorting from template matching, buffering all samples, with 51 percent of area dedicated to memories and 71 percent of power https://ieeexplore.ieee.org/stamp/stamp.jsp?tp=&arnumber=9772722
% 47812 um2 per channel
% 1.74 uW per channel
% Autodencoder 750,000 µm2 per channel, 8.25 mW

\section{Conclusion}

The paper presented a 49-channel neural recording ASIC fabricated in a 28~nm CMOS technology node. This ASIC leverages the sparsity of neural signals, incorporating an event-driven ramp ADC and an on-chip spike compression based on the principal component analysis. The ASIC provides a compression ratio of up to 328. This permits simultaneous transmission of all recording channels while consuming below 1 mW within a compact 1.4~$\times$1.4~\si{\milli\meter\squared} area. This architecture paves the way towards chronically implantable systems with a larger number of dense electrodes, opening up new possibilities for neuroscientific research clinical treatment of neurological disorders.

\section{Acknowledgments}

Authors gratefully acknowledge insightful discussions with Rob Hilkes, Dr Matias Maturana, Professor Steven Prawer and Professor David Garrett. 

% Can use something like this to put references on a page
% by themselves when using endfloat and the captionsoff option.
\ifCLASSOPTIONcaptionsoff
  \newpage
\fi

\bibliographystyle{IEEEtran}
\bibliography{References/bst_control,References/library}

\end{document}